\documentclass[12pt]{article}
\usepackage{graphicx}
\usepackage{amsmath}
\usepackage{amssymb}
\usepackage{bm}

\def\d{\partial}

\def\vk{\bm{k}}
\def\vx{\bm{x}} 
\begin{document}

\title{On quantization of nondispersive wave packets}
\author{M.V.Altaisky\thanks{Space Research Institute RAS, Profsoyuznaya 84/32, Moscow, 117997, 
Russia; and Joint Institute for Nuclear Research, Joliot-Curie 6, Dubna, 141980, Russia; e-mail: altaisky@mx.iki.rssi.ru}, 
N.E.Kaputkina\thanks{National  University of Science and Technology ``MISiS'', Leninsky prospect 4, Moscow, 119049, Russia; e-mail: nataly@misis.ru} \\
}
\date{Aug 21, 2012}
\maketitle
\begin{abstract}
Canonical commutation relations for the Bateman-Hillion type nondispersive wave packets are constructed 
\end{abstract}

\section{Introduction}
The packet-like asymptotic solutions of the wave equation 
\begin{equation}
(\d^2_{x^2} + \d^2_{y^2} +\d^2_{z^2} -\d^2_{t^2})u(x,y,z,t)=0 \label{W:eq}.
\end{equation}
trace their origin from the Bateman's works on conformal symmetry  \cite{Bateman1909,Bateman1955}. Historically such solutions were 
first derived approximately in terms of parabolic equation of the 
diffraction and are related to paraxial optical beams, see \cite{Kiselev2007eng} for a review.
The interest to the localized solutions of wave equation is encouraged 
by the progress in generation of ultra-short laser pulses \cite{BK2000},
in view of the fact that large variety of optical problems can be 
solved in terms of scalar equation.  
The packet-like localized solutions of 
equation \eqref{W:eq}  have a general form \cite{Hillion1992w}: 
\begin{equation}
u=\frac{f(\theta)}{\sqrt{2}\xi_+ -\imath\varepsilon}, \quad \theta = \sqrt{2}\xi_- + \frac{x^2+y^2}{\sqrt{2}\xi_+ - \imath\varepsilon}, \label{bts:eq}
\end{equation}
where $f$ is an arbitrary localized function,
\begin{equation}
\xi_\pm = \frac{z\pm t}{\sqrt{2}} \label{LC:eq},
\end{equation}
are the light-cone coordinates, describing a particle 
moving at a speed of light ($c=1$) along $z$ axis,
$\varepsilon>0$.  
Their particular form with $f(\theta)=e^{\imath q \theta}$ was 
called 'quasiphotons' by Babich and Ulin \cite{BU1982}, or generally 
'focus wave modes' \cite{Brittingham1983}, for the waves which remain 
focused in $\xi_-$ coordinate.

Although the plane-wave solutions $\exp(\imath(\vk\vx-\omega t))$ of equation \eqref{W:eq}
have laid the basis of quantum electrodynamics \cite{Feynman1950,BS1959}, the packet-like solutions of type 
\eqref{bts:eq} were considered only as a special 
case of classical electrodynamics. Their importance has 
raised after the seminal paper of Brittingham \cite{Brittingham1983}, 
who has proved the \eqref{bts:eq}-type 
solutions  -- the focus wave modes -- ({\em i}) 
satisfy the homogeneous Maxwell's
equations, ({\em ii}) are continuous and non-singular, ({\em iii}) have a
three-dimensional pulse structure, ({\em iv}) be non-dispersive for
all time, ({\em v}) move at light velocity in straight lines. This stimulated a series 
of work indented for practical application of localized 
electromagnetic pulses for long-range energy transfer 
without dispersion, electromagnetic and acoustic bullets, 
Gaussian wave beams in optics and geophysics, optical 
information processing in microcavities  \cite{Kiselev1983,MP1990,Overfelt1991,KPC2012}.

The application of the Bateman-Hillion-like solutions or weakly localized wave packets to 
quantum field theory was restricted to the 
formal studies of the localized solutions 
of the Klein-Gordon and the Dirac equations 
\cite{SBZ1990,PF2001eng}. In contrast to the solitons, 
the solutions localized due to nonlinearity, which 
have long been studied in quantum field theory, 
see e.g.\cite{Neveu1977}, neither the Bateman-Hillion  solutions of wave equations nor the Moses-Prosser wave bullets   have been ever 
considered as operator-valued functions and have 
been ever subjected to canonical commutation relations. 

In the present paper we consider the nondispersive solutions 
of wave equation,
as particle-like solutions 
of the field equations, describing a quantum 
particles subjected to canonical commutation relations.
This quantization condition results in certain restrictions on the  
amplitude and the width of the pulse-wave which 
makes it into quantum particle. In the next sections we will consider this 
problem for free scalar field theory models in $1+1$ and $3+1$ dimensions.

\section{$\bm{d=2}$}
The nondispersive localized wave solutions in three plus one dimensions 
are generalizations of the traveling wave solutions of the wave 
equation in one plus one dimension:  
\begin{equation}
(\d^2_{z^2}-\d^2_{t^2})u(z,t)=0 \label{w11:eq},
\end{equation}
which is equivalent to the equation 
$
\d^2_{\xi_+\xi_-} u(\xi_+,\xi_-)=0$. The solution of the equation 
\eqref{w11:eq} is a superposition of two independent 
solutions traveling right and  left along the $z$ axis:
$$
u = f(\xi_-) + g(\xi_+),
$$
where $f$ and $g$ are arbitrary functions. 

In quantum field theory the field $\phi$, which satisfies 
the massless field equation \eqref{w11:eq}, is considered as an operator-valued 
function $\phi = \phi(z,t)$.
Using the Fourier transform the field $\phi$ can be casted as a sum 
of the positive and the negative frequency components 
$$
\phi^{\pm}(x) = \int_{k_0>0} e^{\pm \imath k x} \delta(k^2-m^2) 
\phi(\pm k) dk .
$$
In case of massless field in two dimensions, $x=(z,t)$, this gives 
\begin{equation}
\phi(z,t) = \int \frac{dk}{2\pi 2 \omega_k} \left[ e^{-\imath \omega_k t + \imath k z} \hat u(k) + e^{\imath \omega_k t - \imath k z} \hat{u}^\dagger (k) \right], \label{lf:eq}
\end{equation}
where the integration over $\frac{dkd\omega}{(2\pi)^2}$ is made into 
one dimensional integration $\frac{dk}{2\pi 2 \omega_k}$ using 
the mass-shell delta-function $\delta(\omega^2-k^2)$, which results 
in $\omega_k=|k|$. The operators $\hat u(k)$ and $\hat{u}^\dagger(k)$ are referred 
to as the annihilation and the creation operators for the quanta with momentum $k$.
They satisfy commutation relations 
\begin{equation}
[\hat u(k), \hat{u}^\dagger(k') ] = 2\pi  2\omega_k \delta(k-k') 
\label{cr1}. 
\end{equation}
The equation \eqref{lf:eq} is the basis of field quantization. However, 
its physical interpretation leads to counterintuitive result: if a photon 
is described by plane wave \eqref{lf:eq}, then the absorption of photon 
by photographic plate should expose the whole plate: for the plane wave is present everywhere. In reality the exposure is very local. This prompts us to use some localized function, travelling at a speed of light, instead of plane waves. Same as for classical $c$-valued fields 
the localization is achieved by substituting plane wave by the Fourier 
image of some localized function. This leads to 
the operator-valued solution of the massless field equation 
\begin{equation}
\phi(z,t) = \int \frac{dk}{2\pi 2 \omega_k} \Bigl[ e^{-\imath \omega_k t + \imath k z} c(k)\hat u(k) 
+ e^{\imath \omega_k t - \imath k z} c^*(k) \hat{u}^\dagger (k) \Bigr], \label{llf:eq}
\end{equation} 
here we assume $c^*(k)=c(-k)$.

The canonical momentum, conjugated to the field density $\phi(z,t)$ 
is $\pi(z,t) = \frac{\d\phi}{\d t}$:
\begin{equation}
\pi(z,t) = -\frac{\imath}{2} \int \frac{dk}{2\pi} \Bigl[ e^{-\imath \omega_k t + \imath k z} c(k)\hat u(k) 
- e^{\imath \omega_k t - \imath k z} c^*(k) \hat{u}^\dagger (k) \Bigr]. \label{md:eq}
\end{equation} 

Since we consider the wave packet $\phi$ as a quantum particle we can 
introduce the operator of the "mean field coordinate" $\hat{Q}(t)$ and 
the total momentum $\hat{P}(t)$ of the wave packet
$$
\hat{Q}(t) = \frac{1}{V}\int \phi(z,t) dz, \quad \hat{P}(t) = \int \pi(z',t)dz',
$$
where $V$ is the  volume occupied by the field.
Using the commutation relations \eqref{cr1} for the Fourier modes, 
we get the commutator 
$$
[\hat Q,\hat P] = \frac{\imath}{V} \int dzdz' e^{\imath k(z-z')} |c(k)|^2 
\frac{dk}{2\pi}. 
$$
Thus to ensure canonical commutation relations for the wave packet 
\begin{equation}
[\hat Q,\hat P]=\imath \label{ccr}
\end{equation}
we need to fulfil the constraint 
$$
\int \Lambda(z-z') \frac{dzdz'}{V} = 1,
$$
where 
$$
\Lambda(z) = \int e^{\imath k z} |c(k)|^2 \frac{dk}{2\pi}.
$$
Using the symmetry $|c(k)|^2= |c(-k)|^2$ the constraint that 
ensures canonical commutation relations \eqref{ccr} for the 
operator-valued wave packet \eqref{llf:eq} can be written in the 
form 
\begin{equation}
2\int_0^\infty dz \int_{-\infty}^\infty \cos(kz) |c(k)|^2\frac{dk}{2\pi} = 1.
\end{equation}
For the Gaussian wave packet with $c(k) = A e^{-k^2\sigma^2/2}$ this leads 
to the constraint $A^2=1$ independent on $\sigma$.

\section{$\bm{d=4}$}
Without loss of generality we can consider the Klein-Gordon equation 
in $3+1$ dimensions 
\begin{equation}
(\d^2_{x^2} + \d^2_{y^2} +\d^2_{z^2} -\d^2_{t^2}-m^2)u(x,y,z,t)=0
\label {KG:eq}
\end{equation}
with $m$ set to zero for the case of massless field. The Fourier 
image of the localized solution of the Klein-Gordon equation \cite{PF2001eng} can be written as 
\begin{equation}
g(\sqrt{2}k_+) = \frac{B}{k_+^\delta} \tilde f \left(\frac{k_+}{\sqrt{2}} \right) \delta(k^2-m^2) , \quad k_+ = \frac{k_z+\omega_k}{\sqrt{2}},
\end{equation}
where $B$ is a constant, $\delta=0,\frac{1}{2}$ -- for massless, and massive field, respectively, see \ref{ml:sec}, \ref{apm:sec}.

This leads to 
the operator-valued solution of the  field equation 
\begin{eqnarray}\nonumber 
\phi(\vx,t) &=& \int \frac{d^3\vk}{(2\pi)^3 2 \omega_k} \Bigl[ e^{-\imath \omega_k t + \imath \vk \vx}  g(k_z+\omega_k) \hat{u}(\vk) \\
&+& e^{\imath \omega_k t - \imath \vk \vx} g(-k_z-\omega_k ) \hat{u}^\dagger (\vk) \Bigr], \label{llf4:eq}
\end{eqnarray} 
where the annihilation and the creation operators satisfy the commutation relations 
\begin{equation}
[\hat u(\vk), \hat{u}^\dagger(\vk') ] = (2\pi)^3 \cdot 2\omega_k \cdot \delta(\vk-\vk'), \quad \omega_k = \sqrt{\vk^2+m^2} 
\label{cr4}. 
\end{equation}

The canonical momentum, conjugated to the field density $\phi(\vx,t)$, 
is $\pi(\vx,t) = \frac{\d\phi}{\d t}$:
\begin{eqnarray}\nonumber 
\pi(\vx,t) &=& -\frac{\imath}{2} \int \frac{d^3k}{(2\pi)^3} \Bigl[ e^{-\imath \omega_k t + \imath \vk \vx} g(k_z+\omega_k)\hat u(\vk) \\
&-& e^{\imath \omega_k t - \imath \vk \vx} g(-k_z-\omega_k) \hat{u}^\dagger (\vk) \Bigr]. \label{md3:eq}
\end{eqnarray} 
The "mean field coordinate" $\hat{Q}(t)$ and 
the total momentum $\hat{P}(t)$ of the wave packet are: 
$$
\hat{Q}(t) = \frac{1}{V}\int \phi(\vx,t) d^3\vx, \quad \hat{P}(t) = \int \pi(\vx',t)d^3\vx'.
$$
The constraint \eqref{ccr} results in 
\begin{eqnarray} \nonumber 
1 &=& \int d^3\vx  e^{\imath \vk\vx} g(k_z+\omega_k)g(-k_z-\omega_k) 
\frac{d^3\vk}{(2\pi)^3} \\
 &=& \left. g(k_z+\omega_k)g(-k_z-\omega_k)\right|_{\vk=0}.\label{cc4}
\end{eqnarray}
For Gaussian packet $g(k)=Ae^{-\frac{k^2\sigma^2}{2}}$ this gives 
the normalization constraint  
\begin{equation}
A^2 e^{-\sigma^2 m^2} = 1.
\end{equation}

\section{Conclusions}
 The idea of construction of nondispersive wave packets, which 
 follow a trajectory of classical particle, from the coherent superposition of harmonic oscillators can be traced back to 
 Schr\"odinger \cite{Schrodinger1926}. In classical scales 
 such packets of different form can be created by combination 
 of electromagnetic pulses of equispaced frequencies. In quantum 
 physics, where the energy levels of atoms, used for laser pulse  generation, are not equispaced, the known way to suppress wave packet 
 dispersion is to use external electromagnetic fields \cite{KE1996,BDZ2002}. The most intensive experimental studies have been performed 
 in creation Trojan wave packets -- nondispersive wave packets of electron density on circular orbits \cite{Wyker2011}. The creation of 
 nondispersive packet on a circular orbit is achieved by using circularly polarized laser beams and is a technically sophisticated, 
 however it seems that a simpler problem of creating a nondispersive 
  wave packet moving linearly, which was studied in classical 
 electrodynamics since Brittingham \cite{Brittingham1983} and is implemented experimentally, was 
 not implemented in quantum case, regardless an evident advantage of no need of the external field for suppressing dispersion. In present paper, using a simple scalar model, we constructed the quantization condition for such packets in purely relativistic case. If such mechanism exists in quantum electrodynamics, it may 
 be related to the coherent electromagnetic energy transfer at mesoscopic scales, along with excitons and other nonrelativistic mechanisms, see e.g. \cite{ETC2007}. In future we hope to apply the localized wave packets, subjected to quantization constraints described in this paper, to the scattering problems of localized quantum particles described by quantum field theory methods \cite{AltaiskyPRD10}. In particular the existence of the localized 
focus wave modes \cite{Brittingham1983} in classical electrodynamics may be of interest for the quantum theory of gauge fields \cite{FGL2007}.   
 
\section*{Acknowledgement}
The research was supported in part by the Program of Creation and Development of the National University of Science and Technology "MISiS" and by the RFBR project 11-02-00604a. The authors have benifited from the discussions of these results with Drs. 
A.L.Kataev, A.P.Kiselev, M.V.Perel and O.V.Teryaev.


\appendix
\section{The Bateman-Hillion solution of wave equation \label{bat:sec}}
The Bateman-Hillion solution of the wave equation 
\eqref{W:eq} can be obtained by the change of variable (\S 4.2.1 of \cite{Kiselev2007eng}):
\begin{equation}
u = \int_{-\infty}^\infty \hat{u}e^{\imath \alpha a} da,
\label{bsub}
\end{equation}
where 
$$
\hat{u} = \hat{u}(x,y,\alpha,\beta), \quad \alpha = z-t, \beta = z+t.
$$
The new function $\hat{u}$ satisfies the parabolic equation 
$$
\frac{\d^2 \hat{u}}{\d x^2} + \frac{\d^2 \hat{u}}{\d y^2} 
+ 4\imath a \frac{\d \hat{u}}{\d \beta} = 0,
$$
which has the solution 
\begin{equation}
\hat{u} = \beta^{-1} \exp(\imath a (x^2+y^2)/\beta) \hat{f}(a)
\label{bsol},
\end{equation}
where $\hat{f}(a)$ is arbitrary function. The Bateman-Hillion solution is obtained by substitution of \eqref{bsol} into \eqref{bsub}.

\section{Fourier image of the massless field solution \label{ml:sec}}
To evaluate the Fourier image of the solution \eqref{bts:eq} of wave equation \eqref{W:eq} we use the light-cone variables 
$$ \xi_{\pm} = \frac{z\pm t}{\sqrt{2}}, \quad k_\pm = \frac{k_z\pm\omega}{\sqrt{2}}. $$
In these variables 
\begin{align*}\nonumber 
\tilde{\phi}(k_x,k_y,k_+,k_-) &=& \int 
 dx dy d\xi_- d\xi_+   \frac{f(\theta)}{\sqrt{2}\xi_+} 
e^{-\imath (k_x x + k_y y + k_+  \xi_- + k_- \xi_+)} \\
 &=& \frac{1}{2} \int 
  d\theta d\xi_+ dx dy   \frac{f(\theta)}{\xi_+}  
 \exp\Bigl[ -\imath \bigl(k_x x + k_y y + k_- \xi_+ &+& \frac{k_+}{\sqrt{2}} \left[\theta 
 - \frac{x^2+y^2}{\sqrt{2}\xi_+} \right] \bigr) \Bigr]   
.
\end{align*}
We have the product of two identical integrals in $x$ and $y$ transversal coordinates 
\begin{equation}
\int_{-\infty}^\infty e^{\imath k x^2 - \imath k_x x} dx 
\int_{-\infty}^\infty e^{\imath k y^2 - \imath k_y y} dy = 
\frac{\imath\pi}{k} e^{-\imath \frac{k_\perp^2}{4k}} \label{fi2}
\end{equation}
with $k=\frac{k_+}{2\xi_+}$. The remaining integration in $\xi_+$ is performed by the change of variable $r=\xi_+/k_+$. This gives 
\begin{equation} 
\tilde{\phi}(k_x,k_y,k_+,k_-) = 2\imath \pi^2 \tilde{f} \left( \frac{k_+}{\sqrt{2}}\right) \delta \left(\frac{k_\perp^2}{2} + k_+ k_- \right) 
\label{mlf:eq}
\end{equation}

\section{Fourier image of the massive solution \label{apm:sec}}
To evaluate the Fourier image of the localized solution for the 
Klein-Gordon equation  
we integrate the massless solution in 5d over the extra variable 
$z'$ \cite{PF2001eng}:
\begin{align*}
\tilde{\phi}_m(k_x,k_y,k_+,k_-) = \int e^{-\imath (k_x x + k_y y + m z' + k_+ \xi_- + k_- \xi_+)} \times \\ \nonumber 
 \times \frac{f(\theta)}{(\sqrt{2}\xi_+)^{3/2}} dx dy dz' d\xi_- d\xi_+  
 =  \int e^{-\imath (k_x x + k_y y + m z' + k_- \xi_+)} \times  \\
\times  e^{\imath\frac{k_+}{\sqrt{2}} \left(\theta 
 - \frac{x^2+y^2+{z'}^2}{\sqrt{2}\xi_+} \right) }   
\frac{f(\theta)}{(\sqrt{2}\xi_+)^{3/2}} 
 \frac{d\theta}{\sqrt{2}} d\xi_+ dx dy dz'. 
 \end{align*}
The product of three identical integrals 
in $x,y,z'$, cf. Eq. \ref{fi2}, is equal to 
$$
\frac{(2\pi)^{3/2}}{\sqrt{\imath}}\left(\frac{\xi_+}{k_+} \right)^{3/2}
e^{-\frac{\imath}{2}(k_\perp^2+m^2) \frac{\xi_+}{k_+}}.
$$
Introducing the new variable $r= \frac{\xi_+}{k_+}$ and integrating 
over it we get 
\begin{equation}
\tilde{\phi}_m(k_x,k_y,k_+,k_-) = 
\frac{
\tilde{f}\left( \frac{k_+}{\sqrt{2}} \right) (2\pi)^{5/2}
}{
\sqrt{2\imath k_+} 2^{3/4}
}
\delta \left(\frac{k_\perp^2+m^2}{2}+k_+ k_-\right).
\end{equation}
\end{document}